\documentstyle[prd,aps]{revtex}

\begin{document}

\title{\bf Statistical Gauge Theory for Relativistic Finite Density Problems}

\draft

\author{S. Ying}
\address{Physics Department, Fudan University,
Shanghai 200433, People's Republic of China}
\def\bra#1{\mathopen{\langle#1\,|}}
\def\ket#1{\mathclose{|\,#1\rangle}}

\maketitle
\begin{abstract}

  A relativistic quantum field theory is presented for finite
density problems based on the principle of locality. It is found that,
in addition to the conventional ones, a local approach to the
relativistic quantum field theories at both zero and finite density
consistent with the violation of Bell like inequalities should
contain, and provide solutions to at least three additional problems,
namely, 1) the statistical gauge invariance 2) the dark components of
the local observables and 3) the fermion statistical blocking effects,
base upon an asymptotic non-thermo ensemble. An application to models
are presented to show the importance of the discussions.
\end{abstract}


\section{Introduction}

Besides its relevance to domestic processes like in heavy ion
collisions, in nuclear matter, a consistent theory for relativistic
finite density systems is neither less interesting fundamentally due
to its relevance to profound cosmological questions like the mechanism
for baryogenesis, the nature of dark matter, etc., nor can it be
expected to be trivially derivable from the theoretical framework at
zero density.  

Although such kind of theory, which adopts the basic framework of the
field theoretical representation of non-relativistic many body systems
exists and is widely used in literature, at least two important
qualities that a correct theory for high energy physics should possess
is lacking.  The first one is that the essential ingredient of the
relativistic spacetime, namely the principle of locality, has not been
properly addressed. This is because in the grand canonical ensemble
base on which such a theory is built, the chemical potential is a
global quantity in spacetime.  As it is well known in the studies of
few and many body relativistic systems that it is not enough to
simply adopt the relativistic kinematics for each particles in the
system when the interaction is present.  The less well understood
relative time between particles inside the system exists due to the
lack of a frame independent definition of simultaneity in the
relativistic spacetime which constitutes another fundamental
difference between non-relativistic systems based upon Newtonian
spacetime and the relativistic ones. The second one is related to the
fact that unlike in the non-relativistic condensed matter systems, the
relativistic extended theory has not been systematically tested since
nuclear matter at equilibrium to which this theory should be applied
is hard to prepare domestically to allow such tests. In
addition, question posed for hadronic system are most likely different
from the ones posed for a non-relativistic condensed matter system.

This situation leaves room for new theoretical possibilities to propose
a theoretical framework that can cover not only the
non-relativistic situations, which is a specially limit of the theory
when resolution of observation or energy is low, but also the
relativistic ones based on the principle of locality. While the first
above mentioned quality can be implemented theoretically, as it is
done in the following, the full implementation of the second one is
beyond the scope of a theoretical exploration. But at least a
theoretical exploration can improve the situation by try to answer
questions mostly likely to be experimentally studied in a hadronic
system from domestic to astronomical scale and from zero to finite
density. Here again, as it is shown in the sequel, the principle of
locality plays a key role.

   The global and local measurements in quantum systems is compared in
section \ref{sec:GLM}. The limitation of a global measurement in study
relativistic quantum systems is discussed. A 4-vector field, called
the primary statistical gauge field, is introduced.  An asymptotic
grand canonical ensemble is introduced in section \ref{sec:AGCE}. The
consequence of the statistical gauge invariance is studied and a
semi-quantitative non-perturbative method for study some qualitative
aspects of the problem is introduced. The instability of the asymptotic
grand canonical ensemble is revealed in section \ref{sec:ANTE} where
the asymptotic non-thermo ensemble is shown to be necessary. Section
\ref{sec:Dyn} presents the low energy effective action for the primary
statistical gauge field and its dynamical characteristics. It is shown that
the statistical blocking effects result in the absence of a long range force 
that couples to baryon or nucleon number in the normal
phase of the hadronic matter. A brief summary is given in section
\ref{sec:sum}.

\section{Global and Local Measurements}
\label{sec:GLM}

\subsection{Global Measurements}

The non-relativistic many body theory for a system with variable
particle number is based upon the Grand Canonical Ensemble (GCE)
\cite{Huang}. The partition function for the system in the
GCE is related to its Hamiltonian $\widehat H$ and its particle number
$\widehat N$ in the following way
\begin{equation}
    Z = \lim_{\Omega\to\infty }
           Tr e^{-\beta(\widehat H - \mu \widehat N)}, 
\label{part-func}
\end{equation}
where $\Omega$ is the volume of the system, $\beta = 1/T$ with $T$ the
temperature and $\mu$ is the chemical potential.  The limit
$\Omega\to\infty$ is the thermodynamical limit in which the
equilibrium statistical mechanics describes physical many body
system at equilibrium. A measurement of the ground state
fermion number density in GCE which is given by
\begin{eqnarray}
\overline n &=& \lim_{T\to 0}{\partial\over \partial \mu} \left
({T\over\Omega} \ln Z \right )
\label{gl-rho}
\end{eqnarray}
corresponds to a global measurement. Global measurements are most
frequently carried out in condensed matter systems with a size smaller
than or comparable to our human body (of order 1 meter) which permit us
to study the system ``as a whole''.  The GCE is highly successful in
describing these situations. Questions posed by high-energy physics,
astrophysics and cosmology are somewhat different. The size of the
systems compared to that of the resolution of the
measuring apparatus are normally much
larger, therefore one actually is capable of study either more
details of the probed systems due to the reduction of the wave length
of the probing system or only small fraction of the probed system due
to the impossibility of covering the whole probed system by the
probing system at the {\em same time}. One is in effect doing local
measurements in these later situations.

\subsection{Local Measurements}

 Global measurements are in principle not directly definable since
simultaneity of two events separated from each other by a space-like
distance depends on reference frame in special
relativity. Measurements in relativistic space-time are in principle
local ones in space-time.  The value of a global quantity is obtained
from an integration over the results of a complete set of local
measurements carried out at the same time. In addition, the property
of asymptotic freedom in strong interaction makes the local
measurements relevant to the study of the properties of fundamental
current quarks at short distances where they are point like and almost
free particles.

  Local measurements in quantum field theory are realized by
exerting an external local field to the system at the space-time point
interested, and, the measured quantity are deduced from the response
of the system to the external field. For finite density systems, a
Lorentz 4-vector local field $\mu^\alpha(x)$, called the primary
statistical gauge field, is introduced. 
The ground state expectation value of fermion number density can be
expressed as the functional derivative
\begin{eqnarray}
    \overline \rho(x) &=& {\delta \ln Z\over \delta\mu^0(x)}
\label{lc-rho}
\end{eqnarray}
with $Z$ a functional of $\mu^\alpha$.
It corresponds to a local measurement in the ground state at the
space-time point x.

\subsection{The dark component of local observables}

That $\overline\rho\equiv \overline n$ is not mathematically
warranted. The GCE expression for $\overline n$ in Eq. \ref{gl-rho}
suffers from an ambiguity related to the thermodynamical limit of
$\Omega\to\infty$ which is taken ahead of the partial derivative over
the chemical potential $\mu$ in Eq. \ref{gl-rho}.  Since the
contributing eigenvalues of the number operator $\widehat N$ to
Eq. \ref{gl-rho} are proportional to $\Omega$, the partial derivation
over $\mu$ is not well defined in Eq. \ref{gl-rho} due to the fact
that $\mu$ is multiplied by an infinite number in Eq. \ref{part-func}.
Eq. \ref{lc-rho} does not contain such an ambiguity since the
modification of the system due to a local change in the primary
statistical gauge field $\mu^\alpha(x)$ does not modify the system by
an infinite amount. So Eq. \ref{lc-rho} is an exact expression for the
particle number density for an uniform system.

The difference 
\begin{equation}
\rho_D = \overline\rho - \overline n 
\end{equation}
is called the dark component of the local observable in the ground
state. It is known that $\overline\rho$ equals to $\overline n$ for a
system made of non-interacting fermions and bosons which has a
Hamiltonian (or Lagrangian) quadratic in the fields corresponding to
the particles. This equality is also expected to hold for weakly
interacting systems. As the strength of the interaction or the
non-linearity of the system increases, the dark component $\rho_D$ is
expected to develop non-vanishing value. For a system whose behavior
can be approximately described by an ensemble of quasi-particles, as
it is demonstrated in the following, $\overline n$ contains
contributions from spatially extensive quasi-particles of the system
that can propagate indefinitely in time. So $\rho_D$ contains all
other contributions to $\overline\rho$ due to localized and transient
quantum fluctuations of the fields.

\section{The Asymptotic Grand Canonical Ensemble}
\label{sec:AGCE}

\subsection{The evaluation of conserved local observables}

   Unlike in a weakly interacting theory in which the spectra of the
system are not modified in some essential way from the corresponding
free system, the number (density) of the particle can be found by a
counting of the particles inside the system. The particle number
density in a strongly interacting theory is obscured by the fact that
in certain phases of a strongly interacting system the original
particles are either confined or altered in a fundamental way due to
phase transitions in the system. It makes a straightforward
counting of these particles difficult since some type of them do not
appear in the set of initial and final physical states at all. In
these more general case, one should start from the evaluation of the
original field theoretic definition of the particle number density
operator. For any conserved particle numbers, like the baryon number,
the evaluation can be simplified by counting the particle number of
the system when the strength of the interaction is adiabatically
switched off in the remote past. These conserved numbers are not going
to be changed as the interaction strength is fully switched on long
before the measurement time in the near past and near future.  These
conserved numbers are selected using the asymptotic grand canonical
ensemble (AGCE). It is defined as the grand canonical ensemble of the
corresponding free system in the remote past (or future) when the
interaction effects are adiabatically switched off.

So $\overline\rho $ follows from the asymptotic grand canonical
ensemble (AGCE), which is defined above, with the Lagrangian density
that underlies $Z$ given by
\begin{eqnarray}
    {\cal L}' &=& {1\over 2}\overline\Psi (i\rlap\slash\partial +
      \rlap\slash\hspace{-1pt}\mu O_3 - m)
     \Psi + {\cal L}_B + {\cal L}_{\mbox{\scriptsize int}}
\end{eqnarray}
and $m$ the fermion mass, ${\cal L}_B$ the Lagrangian density of boson
fields of the system and ${\cal L}_{\mbox{\scriptsize int}}$ the
interaction between the boson and fermion fields. An 8 component
``real'' fermion field $\Psi$ is used for the discussion
\cite{8comp-pap} with $O_3$ the third Pauli matrices acting on the
upper and lower 4-component of $\Psi$. In most of the cases, ${\cal
L}_{\mbox{\scriptsize int}}=-\overline\Psi(\Sigma(f)-m)\Psi$ with
$\Sigma$ a function(al) of the boson fields represented by $f$ so that
the fermion degrees of freedom can be integrated out leading to an
effective Euclidean action for the boson fields
\begin{eqnarray}
          S_{\mbox{\scriptsize eff}}[f,\mu] &=& {1\over 2} \mbox{SpLn} {
          [i\rlap\slash\partial +
          \rlap\slash\hspace{-1pt}\mu O_3-\Sigma(f)]\over 
          [i\rlap\slash\partial +
          \rlap\slash\hspace{-1pt}\mu O_3-m]}
           + \int d^4 x \left [{\cal L
          }_B(f) + \mu\overline\rho-\overline e 
          \right ]
\label{eff-action}
\end{eqnarray}
with ``Sp'' the functional trace, $\overline e$ the average energy
density of the corresponding free system of fermions ($\Sigma=0$) and
$\overline\rho$ is the ground state expectation value of fermion
number density to be discussed in the following at the ``chemical
potential'' $\mu\equiv\sqrt{\overline\mu^\alpha \overline\mu_\alpha}$,
where $\overline \mu^\alpha$ is the ground state value of $\mu^\alpha$
normally provided by the external conditions.  The partition
functional
\begin{equation}
Z=\int D[f] exp(S_{\mbox{\scriptsize eff}})
\end{equation}
is then a functional of $\mu^\alpha$ and $\ln Z$ its effective action.

\subsection{The super-selection sector}

The first term in Eq. \ref{eff-action} is invariant under the
$U(1)$ statistical gauge transformation $\mu^\alpha(x)\to
\mu^\alpha(x)-\partial^\alpha\Lambda(x)$, which corresponds to the
conservation of the fermion number.  Since $\mu^\alpha$ is a local
field, its excitation represents certain collective excitations of the
system.  Such a view introduces infinite many extra degrees of freedom
since there is no such a field in the original
theory. These extra degrees of freedom are eliminated by restricting
the representing Hilbert space.

   Superselection-sector in the representing Hilbert space
(SIRHS) containing physical states correspond to the primary
statistical gauge field exists and can be selected using a ``Gauss
Law'' constraint. For the statistical gauge invariant system, it can
be imposed differently on the physical eigenstates of the Hamiltonian
without causing contradictions. States in a SIRHS that is identified by a
coordinate dependent complex function $\varsigma$ satisfy
\begin{eqnarray}
     \bra{\psi^i_\varsigma}
       \left (\widehat \rho + \nabla \cdot \widehat\pi_{u}  \right )
      \ket{\psi^j_\varsigma}
      &=& \delta_{E_i E_j} N_{ij} \varsigma
\label{Q-eigenstate}
\end{eqnarray}
with $\widehat\pi_{u}$ the ``statistical electric field'',
$\ket{\psi^k_\varsigma}$ a physical state that has energy $E_k$,
$\delta_{EE'}$ taking zero or unity value if $E\ne E'$ or $E'=E$
(assuming that $E$ is discrete before length of the time interval in
in which the system is confined is let to go to infinity) and $N_{ij}$
independent of space-time. $\widehat \rho + \nabla \cdot
\widehat\pi_{u}$ is the generator of local statistical gauge
transformations, so the physical states in a SIRHS change a common
(coordinate dependent) phase under a specific gauge transformation
rather than remains invariant The special condition $\varsigma\equiv
0$ is hitherto been used for any gauge field in literature. This is
true for dynamical gauge fields but not necessarily for statistical
gauge fields. Clearly, the later condition is a special case of the
former.

Due to the conservation of the fermion
number, the SIRHS of the system with fixed fermion number selected
\cite{Huang} by the AGCE is invariant during the time evolution. In
the AGCE, the dependence of $\overline\rho$ upon the space-time
independent $\mu$ remains the same whether there is interaction in the
system or not.  It is given by 
\begin{equation}
\overline\rho = N_g \mu^3/3\pi^2
\label{num-density}
\end{equation}
for a massless system, with $N_g$ the total internal degrees of
freedom besides the spin of the fermion. $\overline n$ does not has a
simple dependence on $\mu$, which can be identified with the chemical
potential in uniform case, especially when there is certain kind of
phase transition that modifies the excitation spectra of the system.

In some sense, the AGCE is a GCE for the initial state of the system
under the time evolution, it is a canonical ensemble for the
interaction effects since the invariant SIRHS (under the time
evolution) is fixed in the remote past (or future).  The time
component of the local primary statistical gauge field $\mu^\alpha$
does not necessarily correspond to the chemical potential, which is a
global parameter.  For example, in a system of baryons in thermo
equilibrium, the primary statistical gauge field $\mu^0$ is mainly
non-vanishing around nucleons and nuclei; the chemical potential, on
the other hand, is a constant throughout the system.

\subsection{Relevant results from lattice simulations}

There are interesting lattice Monte-Carlo studies of the finite
density problems \cite{Kogut,Kogut2} based upon GCE for a finite
lattice using Glasgow method.  It is shown using the chiral
Gross-Neveu model that for a given chemical potential, the fermion
number density of the system has two stable values, which are reached
by generating the Monte-Carlo ensemble at either finite or zero
density. The first stable one correspond to the one given by the AGCE
$\rho-\mu$ relation and the second one correspond to the GCE
prediction for a massive system of quasi-particles.  The exact results
for the global treatment of the problem crossover from the later
$\rho-\mu$ curve to the former one near the chiral symmetry
restoration $\mu=\mu_c$. This behavior can be interpreted as that
before the chiral symmetry restoration, the first $\rho-\mu$ curve for
the AGCE is a local minimum of the (Euclidean) action of the various
configurations in the Monte-Carlo ensemble with the GCE curve for the
massive system of quasiparticles the absolute one. Near the chiral
symmetry restoration point, their role exchanges. The existence of a
stable local minimum for the AGCE $\rho-\mu$ curve is a direct
consequence of the conservation of the fermion number in the model
used. It is expected that some of the puzzles encountered in the
lattice gauge theory studies at finite density, like the early on set
of the baryon number before the chiral symmetry restoration, can be
understood in the light of the AGCE since the finite size effects of
the lattice gauge theory study based upon GCE corresponds to, to some
degree, a low resolution local measurement of an infinite system
\cite{dark,8comp-pap}.

\subsection{Semi-quantitative discussions}

This above interpretation for the origin of the dark component of 
$\overline \rho$ can be substantiated by using a cluster
decomposition approximation of the partition functional $Z[\mu]$
\cite{dark} which is expected to provide a non-perturbative picture
for the $\mu$ dependence of the fermion number density at sufficiently
low space-time resolutions. In the crudest approximation, the ground
state fermion number density can be written as \cite{dark}
\begin{eqnarray}
\rho_{\Delta\Omega}(\overline\sigma,\mu)  &=& \sqrt{\alpha\over\pi}
       \int_{-\mu-\overline\sigma}^{\mu-\overline\sigma} d\sigma'
        e^{-\alpha{\sigma'}^2} \rho_\infty(\overline\sigma+\sigma',\mu),
\end{eqnarray}
where $\bar\sigma$ is the ground state expectation value of $\sigma$,
\begin{equation}
\rho_\infty(\sigma,\mu)= {N_g\over 3\pi^2}\left (\mu^2-\sigma^2 \right )^{3/2}
\end{equation}
is the ground state fermion number density in the GCE with fermion
mass $\sigma$. Here $\Delta\Omega$ is the space-time volume that can
be resolved in the observation.  $\alpha$ is proportional to
$\Delta\Omega$ for large enough $\Delta\Omega$; it approaches a
constant value for small $\Delta\Omega$ which scales with the inverse
mass gap of the $\sigma'$ excitation of the system. Clearly,
\begin{equation}
\rho_{\Delta\Omega}(\overline\sigma,\mu)\to \rho_\infty
(\overline\sigma,\mu) = \overline n 
\end{equation}  
as $\Delta\Omega\to\infty$.
This qualitative picture shows that the result of a local measurement
of the fermion number density approaches that of the global one as the
resolution gets lowered; in the limit of the zero resolution, the
result approaches that of the global measurement as expected.

The properties of the fermion propagator can also be discussed using
the cluster decomposition of the partition functional. Following the
same line of reasoning, it can be shown that the fermion propagator
$S_F(x_1,x_2)$ has the following qualitative properties for a
space-like separation between $x_1$ and $x_2$: 1) when the distance
between $x_1$ and $x_2$ is large, $S_F(x_1,x_2)$ behaves much like
that of a genuine particle with mass $\overline \sigma$ 2) when the
distance between $x_1$ and $x_2$ is comparable to the dimension of
$\Delta \Omega$ for which $\alpha$ turns into a constant,
$S_F(x_1,x_2)$ is a superposition of the free fermion propagator with
a mass $\overline\sigma + \sigma'$ weighted by a Gaussian weight
peaked at $\overline\sigma$. In the first case, the propagation of the
fermions inside the system can be well approximated by that of a
quasi-particle. In the second case, the quasi-particle picture for a
fermion is not sufficient due to the fluctuation of
the ``mass term'' can not be suppressed. For a time-like separation,
the qualitative picture discussed above is expected to be still true. 

   The results of non-relativistic statistical mechanics and its
relativistic extension based on GCE are approached by the ones derived
from the local theory under two extreme conditions: 1) vanishing
energy or zero resolution limit and 2) spatial uniformity limit. While
the first limit can be achieved in hadronic processes, the second one
can only be possibly realized in quark-gluon plasma or weak coupling
limit since the confining phase of QCD is not a uniform situation for
the quark degrees of freedom in the view of the local theory even at
equilibrium. This is discussed above. In principle, a constant
chemical potential can only be coupled to the hadronic degrees of
freedom in the confining phase of QCD but not the quark degrees of
freedom. In practice, when the quasi-particle provides a satisfying
set of degrees of freedom in describing the physical problem, an
effective chemical potential for the quasi-particle or constituent
quarks can be used only in an approximate sense in the local theory.

\subsection{The implications}

The existence of the dark component for local fermion number density
has interesting implications. Take the vacuum state of a massless
strongly interacting system in which the chiral symmetry is
spontaneously broken down for example. The originally gapless vacuum
state acquires an excitation gap. If only the massive quasi-particles
are taken into account, the baryon number density are expected to be
non-vanishing only when $\mu$ is larger than the mass of the
quasi-particles. The local fermion number density is however finite as
long as $\mu$ is not zero. A natural question arises as to what this
dark component of fermion number corresponding to?  It is reasonable
to conjecture that this dark component of fermion number corresponds
to those fermion states that are localized and non-propagating
\cite{Stern} similar to the Anderson localization in a condensed
matter system. The random potential here is self-generated within the
system by the transient and localized random quantum fluctuations of
the $\sigma$ and other boson fields. Such a conjecture is amenable to
future studies.

At the fundamental level, the dark component of local observables is
of pure quantum in origin related to the space-like correlations
between local measurements in the relativistic quantum world that are
shown to exist in experimental observations \cite{Aspect,Tapster,Tittel},
which are what reality manifests itself \cite{local}. This is because
the size of the correlated cluster used in studying the vacuum state
(in the Euclidean space-time) tends to zero in the classical limit
($\hbar\to 0$) as a result of the classical relativistic causality,
which suppresses the dark component for any finite resolution
observation \cite{dark}.

\section{The Instability of AGCE and the Asymptotic Non-thermo
         Ensemble}
\label{sec:ANTE}

\subsection{The instability of AGCE}

The Euclidean effective action for the boson field given by
Eq. \ref{eff-action} can be evaluated in the usual way
\cite{8comp-pap}. Since the effective action given by
Eq. \ref{eff-action} is a canonical functional of $\mu^\alpha$, we can
make a Legendre transformation of it, namely 
\begin{equation}
\widetilde
S_{\mbox{\scriptsize eff}} = S_{\mbox{\scriptsize eff}} -\int d^4 x
\overline \mu_\alpha \overline j^\alpha
\end{equation}
with $\overline j^\alpha$ the ground
state current of fermion number density, to make it a canonical
functional of the fermion density in order to study the the stability
of the vacuum state against fluctuations in fermion number. 

{\em The vacuum state is defined as the special ground state that has
the lowest possible ``energy density'' amongst all other ground states,
each one of which has the lowest ``energy density'' under a set of
corresponding external constraints.}

  The effective potential characterizing the above mentioned energy
density can be defined using $\widetilde S_{\mbox{\scriptsize eff}}$
for space-time independent background $f$ fields:
\begin{equation}
V_{\mbox{\scriptsize eff}}=-\widetilde S_{\mbox{\scriptsize
eff}}/V_4
\end{equation}
with $V_4$ the volume of the space-time box that contains
the system.  The stability of the vacuum state against the
fluctuations in fermion number density around $\overline\rho=0 $ can
be studied using $V_{\mbox{\scriptsize eff}}$, which is a canonical
function of $\overline\rho$. Minimization of $V_{\mbox{\scriptsize
eff}} $ with respect to $\mu$ gives the vacuum value $\rho_{vac}$ of
the system since the vacuum $\overline\rho$ is a known function of the
vacuum $\mu$ in AGCE.

The local finite density theory based upon the AGCE developed after
considering the above ingredients is examined by applying it to the
chiral symmetry breaking phase of the half bosonized Nambu
Jona-Lasinio model for the 3+1 dimension and the chiral Gross-Neveu
model for 2+1 dimensions with its effective potential given by
\begin{eqnarray}
V_{\mbox{\scriptsize eff}} &=& - N_g \int {d^Dp\over (2\pi)^D}\left [  \ln
    \left ( 1+{\sigma^2 \over p_+^2} \right )    + \ln
    \left ( 1+{\sigma^2 \over p_-^2} \right ) \right ] \nonumber\\
   &&+ {1\over 4 G_0}
    \sigma^2 + \overline e,
\label{Veff1}
\end{eqnarray}
where ``$D$'' is the space-time dimension, $G_0$ is the coupling
constant and $p^2_\pm = (p_0\pm i\mu)^2+\mbox{\boldmath{p}}^2$.  It
can be shown that $\partial^2 V_{\mbox{\scriptsize eff}}/\partial\mu^2
|_{\mu=0} < 0$ as long as $\sigma\ne 0$ and $D\ge 2$, which implies
that the $\sigma\ne 0$ state with $\overline\rho=0$ is not stable
against fermion number fluctuations. Such a conclusion is both
theoretically and physically unacceptable.

\subsection{The asymptotic non-thermo ensemble}

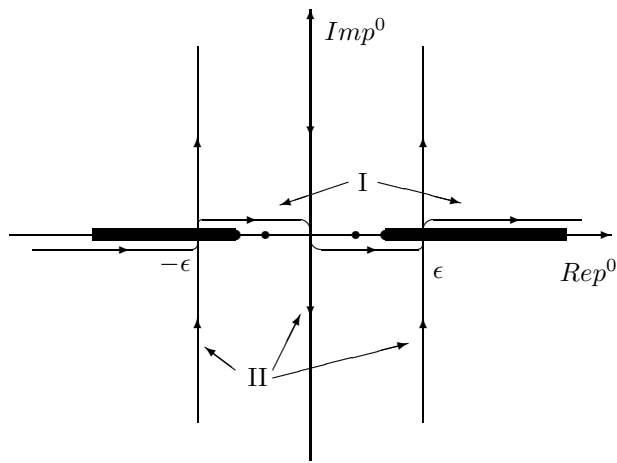
\begin{figure}
\unitlength=1.0mm
\linethickness{0.4pt}
\begin{picture}(120.00,110.00)
\put(40.00,80.00){\vector(1,0){80.00}}
\put(117.00,75.00){\makebox(0,0)[cc]{$Rep^0$}}
\put(86.00,107.00){\makebox(0,0)[cc]{$Imp^0$}}
\put(90.00,79.25){\rule{24.00\unitlength}{1.50\unitlength}}
\put(51.00,79.25){\rule{19.00\unitlength}{1.50\unitlength}}
\put(90.00,80.00){\circle*{1.50}}
\put(70.00,80.00){\circle*{1.50}}
\put(86.00,80.00){\circle*{1.00}}
\put(74.00,80.00){\circle*{1.00}}
\put(43.00,78.00){\line(1,0){20.00}}
\put(97.00,82.00){\line(1,0){19.00}}
\put(67.00,82.00){\line(1,0){11.00}}
\put(82.00,78.00){\line(1,0){11.00}}
\put(63.00,79.00){\oval(4.00,2.00)[rb]}
\put(67.00,81.00){\oval(4.00,2.00)[lt]}
\put(93.00,79.00){\oval(4.00,2.00)[rb]}
\put(97.50,80.50){\oval(5.00,3.00)[lt]}
\put(76.50,80.00){\oval(7.00,4.00)[rt]}
\put(82.00,81.00){\oval(4.00,6.00)[lb]}
\put(53.00,78.00){\vector(1,0){3.00}}
\put(70.00,82.00){\vector(1,0){3.00}}
\put(85.00,78.00){\vector(1,0){4.00}}
\put(102.00,82.00){\vector(1,0){6.00}}
\put(62.00,76.00){\makebox(0,0)[cc]{$-\epsilon$}}
\put(97.00,75.00){\makebox(0,0)[cc]{$\epsilon$}}
\put(80.00,50.00){\vector(0,1){60.00}}
\put(65.00,55.00){\line(0,1){50.00}}
\put(95.00,55.00){\line(0,1){50.00}}
\put(95.00,90.00){\vector(0,1){3.00}}
\put(65.00,90.00){\vector(0,1){3.00}}
\put(95.00,65.00){\vector(0,1){4.00}}
\put(65.00,65.00){\vector(0,1){4.00}}
\put(80.00,95.00){\vector(0,-1){2.00}}
\put(80.00,71.00){\vector(0,-1){2.00}}
\put(73.00,61.00){\makebox(0,0)[cc]{II}}
\put(70.00,62.00){\vector(-4,3){4.00}}
\put(75.00,61.00){\vector(4,1){18.00}}
\put(87.00,87.00){\makebox(0,0)[cc]{I}}
\put(85.00,87.00){\vector(-3,-1){9.00}}
\put(89.00,87.00){\vector(4,-1){11.00}}
\put(75.00,62.00){\vector(1,2){3.67}}
\end{picture}
\caption{\label{fig:contour} 
  The $p^0$ integration contour ${\cal C}$ for the effective
  potential. The thick lines extending to positive and negative
  infinity represent the branch cuts of the logarithmic function.
  Curve ``I'' is the original $p_0$ integration contour for the theory
  in the Minkowski space-time. Curve ``II'' corresponds to the $p_0$
  integration contour for the theory in the Euclidean space-time.}
\end{figure}

It appears that the AGCE is not sufficient for the local finite
density theory, a more general ensemble, called the asymptotic
non-thermo ensemble (ANTE) is introduced to cure this pathology. The
ANTE in the remote past of the system is not necessarily a strict
thermal one.  This is because the $\sigma\ne 0$ phase, which is called
the $\alpha$-phase of the massless quark system, is known to be
condensed with macroscopic number of bare fermion-antifermion
pairs. These pairs occupy the low energy states of the system; they
block other fermions from further filling of these states. This kind
of statistical blocking effect is not encoded into Eq. \ref{Veff1},
which starts the time evolution of the system from a set of initial
states having zero number of fermion-antifermion pairs. The later
initial states do not overlap with the states having a finite {\em
density} of such pairs in the thermodynamic limit. Its effects can
however be included in the boundary condition for the system like what
has been done for the finite density cases. Therefore, it is proposed
that the initial ensemble of quantum states in the ANTE for the system
at zero temperature and density are those ones with negative energy
states filled up to $E=-\epsilon$ rather than $E=0$ and with positive
energy states filled up to $E=\epsilon$ rather than empty. It is
expected that such a state can has sufficient overlap with the true
vacuum state of the system for properly determined $\epsilon$. In the
ANTE \cite{8comp-pap}, the effective potential corresponding to
Eq. \ref{Veff1} is expressed as
\begin{eqnarray}
V_{\mbox{\scriptsize eff}} &=& - N_g \int_{\cal C} 
   {d^Dp\over (2\pi)^D}\left [  \ln
    \left ( 1+{\sigma^2 \over p_+^2} \right )    + \ln
    \left ( 1+{\sigma^2 \over p_-^2} \right ) \right ] \nonumber\\
   &&+ {1\over 4 G_0}
    \sigma^2 + \overline e_{(+)} + \overline e_{(-)} - \overline e,
\label{Veff2}
\end{eqnarray}
where 
\begin{equation}
\overline e_{(\pm)} = 2 N_g \int^{\mu_{\pm}} {d^{D-1}
p\over(2\pi)^{D-1}} p.
\end{equation}
Here the upper boundary of the radial $p$
integration in the D-1 dimensional momentum space is denoted as
\begin{equation}
\mu_{\pm} \equiv \mu\pm \epsilon.
\end{equation}
The contour ${\cal C}$ for the (complex) $p_0$ integration is shown in
Fig. \ref{fig:contour} in which both the original Minkowski contour
and its Euclidean distorted contour are drawn.  The effective potential for the
$\sigma\ne 0$ case in ANTE given by the above equation has two sets of
minima, the first set contains $\mu=\pm\mu_{vac}$ and $\epsilon=0$
points, the second one includes $\mu=0$ and $\epsilon= \pm
\epsilon_{vac}$ ones with finite $\mu_{vac}$ and $\epsilon_{vac}$. The
second solutions correspond to the absolute minima, namely, the vacuum
state. Thus there is no instability against quantum fluctuation over
$\rho$.

\subsection{Model studies}

The vacuum and ground states of the strong interaction could have
different phases from the $\alpha$-phase
\cite{lettB,annPhy,Paps-a,Paps-b,Paps-c}. They are characterized by a
condensation of diquarks. Such a possibility is interesting because it
may be realized in the early universe, in astronomical objects and
events, in heavy ion collisions, inside nucleons \cite{Paps-b,nuc-pap}
and nuclei, etc.. For the possible scalar diquark condensation in the
vacuum, a half bosonized model Lagrangian is introduced
\cite{Model-I,8comp-pap}, which reads
\begin{eqnarray}
 {\cal L}_I & = & {1\over 2} \overline \Psi\left
             [i{\rlap\slash\partial}-\sigma- i\vec{\pi}\cdot
             \vec{\tau}\gamma^5 O_3-\gamma^5 {\cal A}_c\chi^c
             O_{(+)}-\gamma^5 {\cal A}^c\overline\chi_c O_{(-)} \right
             ]\Psi -{1\over 4 G_0} (\sigma^2 + \vec{\pi}^2) + {1\over
             2 G_{3'}} \overline\chi_c \chi^c ,\label{Model-L-1}
\end{eqnarray}
where $\sigma$, $\vec{\pi}$, $\overline\chi_c$ and $\chi^c$ are
auxiliary fields with $(\chi^c)^{\dagger} = - \overline\chi_c$ and
$G_0$, $G_{3'}$ are coupling constants of the model.  ${\cal A}_c$ and
${\cal A}^c$ $(c=1,2,3)$ act on the color space of the quark; they are
${\cal A}_{c_1c_2}^c = -\epsilon^{cc_1c_2}$ ${\cal A}_{c,c_1c_2} =
\epsilon^{cc_1c_2}$ with $\epsilon^{abc}$ ($a,b = 1,2,3$) the total
antisymmetric Levi--Civit\'a tensor. Here $O_{(\pm)}$ are raising and
lowering operators respectively in the upper and lowering 4 components
of $\Psi$.

This model has two non-trivial phases. The vacuum expectation of
$\sigma$ is non-vanishing with vanishing $\chi^2 \equiv-
\overline\chi_c \chi^c$ in the $\alpha$-phase. The vacuum state in the
$\alpha$-phase is condensed with quark-antiquark pairs. The vacuum
expectation of $\sigma$ is zero with finite $\chi^2$ that
spontaneously breaks the $U(1)$ statistical gauge symmetry in the
second phase, which is called the $\omega$-phase. There is a
condensation of correlated scalar diquarks and antidiquarks
in the color $\overline 3$ and $3$ states in the $\omega$-phase of the
vacuum state.

Since diquarks and antidiquarks are condensed in the
$\omega$-phase, it is expected that an exchange of the role of
$\epsilon$ and $\mu$ occurs. It is found to be indeed true: there
are also two sets of minima for the effective potential, the first set
is the one with finite $\mu=\pm\mu_{vac}$ and $\epsilon=0$ and the
second set corresponds to $\mu=0$ and finite
$\epsilon=\pm\epsilon_{vac}$ in the $\omega$-phase of the model. But
here the absolute minima of the system in the $\omega$-phase
correspond to the first set of solutions, in which the CP/T invariance
and baryon number conservation are spontaneously violated due to the
presence of a finite vacuum $\mu^\alpha_{vac}$. This conclusion is
also applicable to the $\beta$-phase of models \cite{lettB,annPhy}
with vector fermion pair and antifermion pair condensation, in which
the chiral symmetry $SU(2)_L\times SU(2)_R$ is also spontaneously
broken down.

\subsection{The implications}

   Although it is found that $ \epsilon_{vac} < \sigma_{vac}$ in all
cases that were studied, the effects of a finite vacuum $\epsilon$ can
still have dynamical consequences \cite{8comp-pap,dark,pap2} since at
short distances (between $x_1$ and $x_2$), the propagation of the
fermion is not described by the quasi-particle propagator which has a
definite mass ($\sigma_{vac}$ in the half-bosonized four fermion
interaction model). The fluctuation in the effective quasi-particle
mass are finite at short distances so that there will be a finite tail
of the mass term that goes below $\epsilon_{vac}$ which causes real
dynamical effects some of which are discussed in Refs. \cite{pap2}. 

   Quasi-particles in an interacting system can not propagate
indefinitely so that the fluctuation in their effective mass term can
be ignored. The quasi-particles in the system are scattered constantly
when they propagate inside the system with a finite time to go as a
free particle before being scattered. Therefore the presence of the
statistical blocking effects characterized by a finite $\epsilon$ has
observable effects in principle. One of them are discussed in the
following related to the absence of additional long range force \cite{fifthf}
to the
gravity in the $\alpha$-phase, which is believed to be the phase in
which the strong interaction vacuum state is at at the present day
conditions \footnote{So far, the search for the ``fifth force'' has
been turning out null results.}.

\section{The Dynamics of the Primary Statistical Gauge Field}
\label{sec:Dyn}

The equilibrium configuration of the primary statistical gauge field
$\mu^\alpha$ can has non-trivial topology \cite{8comp-pap} which will
be studied in other works. The vibration of the primary
statistical gauge field $\mu^\alpha(x)$ around its equilibrium
configuration represents certain collective excitations of the
system. 

The dynamics of the primary statistical gauge field are generated by
$S^{(\mu)}_{\mbox{\scriptsize eff}}[\mu]=\ln Z$. Due to the
statistical gauge invariance, the leading order in the derivative
expansion of $S^{(\mu)}_{\mbox{\scriptsize eff}}[\mu]$ in terms of
$\mu^\alpha$ can be expressed as the following: in the normal phase,
the effective action for slow varying
$\mu'_\alpha=\mu_\alpha-\overline\mu_\alpha $ is
\begin{eqnarray}
      S^{(\mu)}_{\mbox{\scriptsize eff}} 
       &=& \int d^4 x \left [ -{Z^{(\mu)}\over 4} f_{\mu\nu}
      f^{\mu\nu} +  {N_g\over \pi^2} {\overline \epsilon}^2 
       \mu' \cdot \mu' \right ] + \ldots 
\label{Seff-mu2}
\end{eqnarray}
with $f^{\alpha\beta} = \partial^\alpha
{\mu'}^\beta-\partial^\beta{\mu'}^\alpha$ and $\overline\epsilon$
the ground state value of $\epsilon$; in the phase where
local statistical $U(1)$ gauge symmetry is spontaneously broken down
and before considering electromagnetic interaction,
\begin{eqnarray}
      S^{(\mu)}_{\mbox{\scriptsize eff}} &=& {1\over 2}
         \int d^4 x \left [
             {i}\overline j^\alpha \overline j^\beta
       + g^{\alpha\beta} {\Pi^{(\mu)}}
       \right ] \mu'_\alpha
      \mu'_\beta + \ldots.
\label{Seff-mu3}
\end{eqnarray}
In both of the situations with slow varying $\mu'_\alpha$, $Z^{(\mu)}$
and $\Pi^{(\mu)}$, which can be extracted from the time-ordered
current--current correlator $< \mbox{T} j^\alpha(x) j^\beta(x')>$, are
approximately constants.

The excitation corresponds to $\mu^\alpha$ is massive for the vacuum
state in the $\alpha$-phase due to the statistical blocking effects.
There is no long range force associated with $\mu^\alpha$ in the
$\alpha$-phase. This conclusion is important for the current theory to
be consistant with empirical facts \cite{fifthf} related to the
absence of long range force between charge neutral objects besides the
gravity. This is because the primary statistical gauge field
$\mu^\alpha$ is massless due to the statistical gauge invariance
before considering the effects of the statistical blocking. The
exchange of massless objects generates long range force between the
interacting objects.

 It can be shown \cite{8comp-pap} that after including
the electromagnetic interaction, the spatial component of $\mu^\alpha$
excitation is massless for the vacuum state in the $\beta$- and
$\omega$- phases in the rest frame of matter.

\section{Summary}
\label{sec:sum}

In summary, it is found that general local relativistic quantum field
theory (in the sense of including both the zero and the finite density
situations) contains dark components for local observables,
statistical gauge invariance and ferminic blocking effects.  The
proper boundary condition for a consistent theory is that of an
optimal ANTE. The rich implications of the finding presented here
remain to be explored.

\section*{Acknowledgement}

   This work is supported by the National Natural Science Foundation
of China and Department of Education of China.

\end{document}